\documentstyle[12pt,amsfonts]{article}
\def\one{1\hskip-.37em 1}
\def\op{{\overrightarrow p}}
\def\oq{{\overrightarrow q}}
\def\oP{{\overrightarrow P}}
\def\oQ{{\overrightarrow Q}}
\def\ox{{\overrightarrow x}}
\def\oy{{\overrightarrow y}}
\def\oR{{\overrightarrow R}}
\def\oS{{\overrightarrow S}}

\def\half{\textstyle{\frac{1}{2}}}
\def\quarter{\textstyle{\frac{1}{4}}}

\def\H{{\cal H}}

\def\ep{\epsilon}

\def\H{{\cal H}}

\def\l{\lambda}

\def\ra{\rightarrow}
\def\tint{{\textstyle\int}}

\def\s{\hskip.08em}

\def\d{\partial}
\def\o{\overline}

\def\b{\begin{eqnarray*}}  
\def\e{\end{eqnarray*}}    
\def\bn{\begin{eqnarray}}  
\def\en{\end{eqnarray}}   
\def\<{\langle}
\def\>{\rangle}

\def\{{\lbrace}
\def\}{\rbrace}
\begin{document}
\title{On the role of coherent states \\in quantum foundations}
\author{John R. Klauder
\\Department of Physics and \\Department of Mathematics\\
University of Florida\\
Gainesville, FL 32611-8440\\
Email: klauder@phys.ufl.edu}
\date{ }
\bibliographystyle{unsrt}
\maketitle
\begin{abstract}
Coherent states, and the Hilbert space representations they generate,
provide ideal tools to discuss classical/quantum relationships. In this
paper we analyze three separate classical/quantum problems using
coherent states, and show that useful connections arise among them. The
topics discussed are: (1) a truly natural formulation of phase space
path integrals; (2) how this analysis implies that the usual classical formalism is
``simply a subset'' of the quantum formalism, and thus demonstrates a universal
coexistence of both the classical and quantum formalisms; and (3) how these
two insights lead to a complete analytic solution of a formerly insoluble family
of nonlinear quantum field theory models.

\hskip5.5em PACS  ~~03.65 Ca, 03.65 Db, 03.65 Ta
\end{abstract}
Old wine comes in old bottles, and some of that wine merits the
complimentary phrase: vintage wine. So too is it
with old wisdom that flows from the use of old coherent states. This paper discusses
how coherent states---very much like those commonly used in the study of quantum
optics---have a great deal to teach us about the classical/quantum connection. Here,
we offer three vintage stories, which also build on one another, demonstrating how coherent states can illuminate and clarify issues that
are basic to an improved understanding of the classical/quantum connection.

We start with the first story.

\subsection*{Can Phase Space Path Integrals be Canonically \\Covariant Under Coordinate Transformations?}
The basic Hermitian operators of a single quantum degree of freedom, namely, $Q$ and $P$, satisfy the
commutation relation $[Q,\s P]=i\s\hbar\s\one$, and
posses a complete set of (formal) eigenvectors and eigenvalues of the form
$Q\s|q\> =q\s |q\>$, $q\in{\mathbb R}$, and $P\s|p\> =p\s|p\>$, $p\in{\mathbb R}$,
with $\<q|q'\>=\delta(q-q')$, $\<p|p'\>=\delta(p-p')$, and $\<q|p\>=e^{iqp/\hbar}/\sqrt{2\pi\hbar}$.
   These very relations imply the traditional resolution of unity in the form
    $\tint |q\>\s\<q|\,dq = \tint |p\>\s\<p|\,dp = \one$.
   The set of states $\{|q\>\}$ and $\{|p\>\}$ lead to functional representations of an abstract Hilbert space
   via the functions $\psi(q)\equiv\<q|\psi\>$ and (with a traditional abuse of notation)
   $\psi(p)\equiv\<p|\psi\>$;
   in turn, the inner product of two abstract vectors is realized as
     $\<\psi|\phi\>=\tint \psi(q)^*\phi(q)\,dq=\tint \psi(p)^*\phi(p)\,dp$.
     The physical meaning of $\psi(q)$ arises when
        $\tint |\psi(q)|^2\,dq=\tint |\psi(p)|^2\,dp=1$,
     which normalizes the abstract vector so that $\|\s|\psi\>\s\|\equiv+\sqrt{\<\psi|\psi\>}=1$. In that
     case, $|\psi(q)|^2$ denotes the probability density to find the particle at position $q$ and
     $|\psi(p)|^2$ denotes the probability density to find the particle at momentum $p$ (at least with
     the usual meaning of the variables $q$ and $p$). This physical interpretation strongly depends on the
     fact that the state $|q'\>$ is orthogonal to $|q\>$ whenever $q'\not= q$; and similarly for $p'$ and $p$.
     In that case $\psi(q')$ is independently specifiable from $\psi(q)$, a necessary requirement for the
     given probability interpretation.

     The propagator $K(q'',T;q',0)$ transports the state from time $t=0$ to $t=T$ as
        $\psi(q'',T)=\tint K(q'',T;q',0)\,\psi(q',0)\,dq'$, and the propagator follows [with $\hbar=1$ and $\ep=T/(N+1)$] from
      \b && \hskip-2em K(q'',T;q',0)=\<q''|\s e^{ -i\H\s T}\s|q'\>\\
         &&=\int \<q''|\s e^{ -i\H\s \ep}\s|q_N\>\<q_N|\s e^{ -i\H\s \ep}\s|q_{N-1}\>\cdots\<q_1|\s e^{ -i\H\s \ep}\s|q'\>\,\Pi_{n=1}^N\,dq_n\\
         &&=\int \<q''|p_{N+1/2}\>\<p_{N+1/2}|\s e^{ -i\H\s \ep}\s|q_N\>\<q_N|p_{N-1/2}\>\<p_{N-1/2}|\s e^{ -i\H\s \ep}\s|q_{N-1}\>\\
         &&\hskip3em\times\cdots\<q_1|p_{1/2}\>\<p_{1/2}|\s e^{ -i\H\s \ep}\s|q'\>\,
         \Pi_{n=0}^N\,dp_{n+1/2}\,\Pi_{n=1}^N\,dq_n\;.\e
         For very small $\ep$, the matrix element involving the Hamiltonian can be approximated to first order by
           $ \<p|\s ^{ -i\H\s\ep}\s|q\>\simeq \<p|\s(1-i\H\s\ep\s)\s|q\>\simeq\<p|q\>\,e^{-i\ep H(p\s;\s q)}$,
           where $H(p\s;q)\equiv\<p|\s\H\s|q\>/\<p|q\>$. Putting this together, it follows that
           \b &&\hskip-2em K(q'',T;q',0)=
           \lim_{N\ra\infty}\,(2\pi\hbar)^{-(N+1)}\,\int e^{ (i/\hbar)\Sigma_{n=0}^N
           [p_{n+1/2}(q_{n+1}-q_n)-\ep\s H(p_{n_1/2};\s q_n)\s]}\\
           &&\hskip16em\times\Pi_{n=0}^N\,dp_{n+1/2}\,\Pi_{n=1}^N\,dq_n\;,\e
           where the limit is taken so that $\ep(N+1)=T$. Interchanging the limit and the integrations---an
           operation that is {\it not} mathematically justified---and writing for the integrand the form it takes for continuous
           and differentiable paths, leads to the usual formal expression, namely,
            \b K(q'',T;q',0)={\cal N}\int e^{ (i/\hbar)\tint[\s p\s{\dot q}-H(p,q)\s]\,dt}\,{\cal D}p\,{\cal D}q\e
            with the identification that $H(p,q)=H(p\s;q)$. [Other variations on this theme can be
            developed that lead to manifestly real expressions for $H(p,q)$.]

            It is frequently observed that in the limit in which $\hbar\ra0$, a stationary phase evaluation of the
            phase space path integral asserts that the principal contribution to the path integral arises
            from stationary points, namely, where, to first order,
              $ \delta\tint[\s p\s{\dot q}-H(p,q)\s]\,dt=0$.
              This  is just Hamilton's variational principle to derive the classical equations of motion,
              and thus this argument exhibits how the classical theory emerges from the quantum theory. But while this construction of the phase space path integral is correct, it is {\it not} natural since the ``classical paths'' have alternating times where $q$ is known exactly---and thus $p$ is {\it un}known---and vice versa. This is simply not natural!   Moreover, the
              formal path integral appears to be covariant under classical canonical
              coordinate transformations, but that view is countered by the fact that, despite a suggestive similarity to Liouville's Theorem, the formal measure
              ${\cal D}p\,{\cal D}q$, is {\it not} invariant under canonical coordinate transformations
              since the derivation of the path integral clearly shows there is one more $p$ integration
              than $q$ integration. Thus this formulation of the path integral cannot be covariant under canonical transformations.

              We take another look at this story from the point of view of coherent states \cite{kla1}.
              Coherent states, defined for our purposes as $|p,q\>\equiv e^{ -iqP/\hbar}\s e^{ipQ/\hbar}\s|0\>$, for all $(p,q)\in{\mathbb R}^2$, where $|0\>$ is a unit vector that satisfies $(Q+iP)\s|0\>=0$, also admit a resolution of unity in the usual form $\tint|p,q\>\<p,q|\s d\mu(p,q)=\one$, where $d\mu(p,q)=dp\s dq/2\pi\hbar$. The coherent states are {\it not} mutually orthogonal; instead, the overlap function reads $\<p,q|p',q'\>=\exp\{i(p+p')(q-q')/2\hbar-[(p-p')^2+(q-q')^2]/4\hbar\}$. Nevertheless, such states admit a functional representation of an abstract Hilbert space via $\psi(p,q)=\<p,q|\psi\>$, with an inner product given by $\<\psi|\phi\>=\tint
              \psi(p,q)^*\phi(p,q)\s d\mu(p,q)$. The physical interpretation of $p$ and $q$ for coherent states
              follows from the fact that $\<p,q|P|p,q\>=p$ and $\<p,q|Q|p,q\>=q$, namely, they are {\it mean values} in the coherent states.

              The propagator in the coherent state representation is given [with $(p'',q'')=(p_{N+1},q_{N+1})$, $(p',q')=(p_0,q_0)$ , and $\hbar=1$] by \cite{kla2}
              \b && K(p'',q'',T;p',q',0)=\<p'',q''|\s e^{-i\H\s T}\s|p',q'\>\\
                &&=\int \Pi_{n=0}^N\<p_{n+1},q_{n+1}|e^{-i\ep\H}|p_n,q_n\>\,\Pi_{n=1}^N\,d\mu(p_n,q_n)\\
                &&=\lim_{N\ra\infty}\int \exp(\Sigma_{n=0}^N\{i\s\half(p_{n+1}+p_n)(q_{n+1}-q_n)\\
                &&\hskip2em-\quarter[(p_{n+1}-p_n)^2+(q_{n+1}-q_n)^2]
                -i\ep\s H(p_{n+1},q_{n+1};p_n,q_n)\}\s)\\
                &&\hskip4em\times\Pi_{n=1}^N\s d\mu(p_n,q_n)\;, \e
                where $H(p,q;p',q')\equiv \<p,q|\H|p',q'\>/\<p,q|p',q'\>$. Again a formal (and unallowed)
                interchange of limit and integrations leads to the formal expression
                  \b K(p'',q'',T;p',q',0)={\cal M}\int e^{(i/\hbar)\tint[p\s{\dot q}-H(p,q)]\,dt}\,{\cal D}p\,
                  {\cal D}q\;,\e
                  where the symbol $H(p,q)=\<p,q|\H|p,q\>$, i.e., $\H$ is normal ordered, $\H= \s :H(P,Q):$. Again a stationary phase argument would lead to Hamilton's equations, but now the solution is {\it natural} since
                  as expectation values both $p$ and $q$ can be simultaneously specified for all $t\s$!

                  An alternative regularization process is available for coherent state path integrals that involves
                  the insertion of a regularization factor into the formal phase space path integral so that
                  \cite{dkla}
                  \b  K(p'',q'',T;p',q',0)\hskip-1.4em
                     &&=\lim_{\nu\ra\infty}{\cal N_\nu}\int e^{(i/\hbar)\tint[p\s{\dot q}
                     -H(p,q)]\,dt}\,e^{-(1/2\nu)\tint[{\dot p}^2+{\dot q}^2]\,dt}\,{\cal D}p\,{\cal D}q\\
                      &&=\lim_{\nu\ra\infty}(2\pi\hbar)\s e^{\nu T/2}\int e^{(i/\hbar)\tint[ p\s dq-H(p,q)\s dt]} \,d\mu^\nu_W(p,q)\;, \e
                      where in the last line we have introduced a two-dimensional Wiener measure on a flat plane using Cartesian coordinates; this Wiener measure is pinned so that $(p(T),q(T))=(p'',q'')$ and
                       $(p(0),q(0))=(p',q')$. The integral $\tint p\s\s dq$ is a Stratonovich (mid-point rule) stochastic integral that obeys the ordinary rules of calculus, and in this case, the symbol
                       $H(p,q)$ is anti-normal ordered. Observe that the last expression is a genuine path integral composed of continuous paths $p(t)$ and $q(t)$ for any finite $\nu\s$!

                    A canonical transformation is an invertible map from $(p,q)$ to $({\o p},{\o q})$ such that
                    $p\s\s dq={\o p}\s\s d{\o q}+d{\o G}({\o p},{\o q})$.
                      We define coherent states as scalars so that $|{\o p},{\o q}\>=
                   |p,q\>$, which means that the coherent state propagator transforms as a scalar and becomes
                     \b {\o K}({\o p}'',{\o q}'',T;{\o p}',{\o q}',0)
                     =\lim_{\nu\ra\infty}(2\pi\hbar)\s e^{\nu T/2}\s\int e^{(i/\hbar)\tint[{\o p}\s d{\o q}+d{\o G}({\o p},{\o q})-{\o H}({\o p},{\o q})\s dt]}\,d{\o \mu}^\nu_W({\o p},{\o q})\,,\e
                     where ${\o H}({\o p},{\o q})=H(p,q)$, and
                     the Brownian motion is still on a plane but, generally, it is expressed in curvilinear coordinates. This is the only formulation of quantum mechanics that is genuinely
                     covariant under arbitrary canonical coordinate transformation known to the author. Thus,
                     the answer to the first question is yes!

                     \subsection*{Can this formalism help bring classical and quantum \\theories into a single unified picture?}
                    The action functions for quantum and classical mechanics are given by
                       \b I_{quantum}&&\hskip-1.4em=\tint \<\psi|(i\hbar\d/\d t-\H)|\psi\>\,dt\;,\\
                          I_{classical}&&\hskip-1.4em=\tint[p\s{\dot q}-H(p,q)]\,dt\;. \e
                     Varying the first equation leads to Schr\"odinger's equation for $|\psi(t)\>$;
                     varying the second equation leads to Hamilton's equations for $p(t)$ and $q(t)$.
                     These are two vastly different realms, or so it would seem.

                    Suppose we now {\it restrict} the variations that are allowed  for the quantum action.
                    In particular, consider limiting the set of states $|\psi(t)\>$ to just coherent states,
                    i.e., $|\psi(t)\>=|p(t),q(t)\>$, in which case an elementary calculation shows that
                      \b I^{restricted}_{quantum}=\tint\<p,q|(i\hbar\d/\d t-\H)|p,q\>\,dt =\tint[p\s{\dot q}-H(p,q)]\,dt\;, \e
                      where $H(p,q)\equiv\<p,q|\H|p,q\>$ as we have seen before. The result is exactly the classical
                      action, and the restricted quantum equations of motion are just those of the
                      classical realm. Such a restriction is entirely reasonable since {\it macro}scopic experiments
                      on a {\it micro}scopic system are limited to changing its position and its velocity. An important difference between the two views of ``classical theory'' is that in the usual classical theory the physical meaning of $p$ and $q$ is the exact momentum and position of the point particle, while for the restricted quantum approach the physical meaning of $p$ and $q$
                      is as mean values of the momentum and position. Apart from possible $O(\hbar)$ terms in the Hamiltonian, this change of physical interpretation of $p$ and $q$ is the
                      only distinction, and after all, it is only an idealization to say that in the classical theory $p$ and $q$ are ``exact values''; nobody ever measures such quantities to $10^{137}$ decimal places to verify that presumed idealization! The conclusion, therefore, is that classical
                      mechanics is nothing but a {\it subset} of quantum mechanics. And the process of quantization
                      is simply the process of enlarging the domain of definition for the restricted quantum action
                      function \cite{kla4}. From this viewpoint, therefore, there is only one action principle for physics, that of the quantum theory. Thus the answer to the second question is also yes!

\subsection*{Can this single unified picture help solve a \\particular insoluble quantum field problem?}
     Consider the classical Hamiltonian for a many degree of freedom problem given by
          \b H(\op,\oq)=\half [\op^2 +m^2_0\oq^2]+ \lambda_0(\oq^2)^2\;, \e
          where $\op^2\equiv\Sigma_{n=1}^N \s p_n^2$ and $\oq^2\equiv\Sigma_{n=1}^N \s q_n^2$, and $N\le\infty$. When $N<\infty$,
          there is no restriction on the values of $\op$ and $\oq$; when $N=\infty$, the values of $p_n$ and $q_n$ must decrease as $n$ becomes large so that $\op^2+\oq^2<\infty$. These models
          have a rotational symmetry that is helpful in their analysis; indeed, the models in question are called ``Rotationally Symmetric'' models \cite{kla5,akt}. Given $\op$ and $\oq$, there are three rotationally invariant variables: $X\equiv \op^2$, $Y\equiv\op\cdot\oq$, and
          $Z\equiv\oq^2$. The energy is $E=\half X+\half m^2_0 Z + \l_0 Z^2$, and the magnitude of the angular momentum is given by $\overrightarrow{L}^2=(\op\times\oq)^2=X\s Z-Y^2\ge0$.
          The solution to the equations of motion is equivalent to a {\it one} dimensional problem ($N=1$) if
          $\overrightarrow{L}=0$; the solution is equivalent to a {\it two} dimensional problem ($N=2$) if $\overrightarrow{L}\not=0$.

          Quantum mechanically, the variables $\op\ra\oP$ and $\oq\ra\oQ$ such that $[Q_j,P_k]=i
          \hbar\delta_{j, k}\one$. The story is very different for  $N<\infty$ and for $N=\infty$. First,
          for $N<\infty$, the Hamiltonian is given by
            \b \H=\half \oP^2+\half m^2_0\oQ^2 +\l_0 (\oQ^2)^2\;. \e
            In an effort to minimize difficulties that emerge as $N\ra\infty$, we could also choose
            \b \H=\half:\oP^2:+\half m^2_0:\oQ^2:+\l_0:(\oQ^2)^2:\;, \e
            where $:~:$ denotes normal ordering. In the latter case we would have
               \b H(\op,\oq)\equiv\<\op,\oq|\H|\op,\oq\>=\half \op^2+\half m^2_0\oq^2+\l_0(\oq^2)^2\;, \e
               where the coherent states are given by $|\op,\oq\>=\exp(-i\oq\cdot\oP/\hbar)\s \exp(i\op\cdot\oQ/\hbar)\s|0\>\equiv U[\op,\oq]\s|0\>$.
               The normalized fiducial vector $|0\>$  satisfies $(m_0\oQ+i\oP)\s|0\>=0$ or equivalently
               $(m_0\oQ-i\oP)\s(m_0\oQ+i\oP)\s|0\>=\s\s\s:\oP^2+m^2_0\oQ^2:\s|0\>=0$. However, this construction
               works only when $N<\infty$. As $N\ra\infty$, the operator $:\oQ^2:$ diverges proportional to
               $N$, and it must be regularized by dividing by $N$. However, $:\oQ^2:/N$ converges to a multiple
               of the identity operator. In order to obtain a valid limit as $N\ra\infty$, it is necessary that $\l_0$ is replaced by
               $\l_0/N$ (to ensure the regularization of the nonlinear term), and the result converges to a free field with a different mass value. This is hardly a proper quantum solution to the original nonlinear classical model!

               For $\l_0=0$, the free theory, the operator $\H_0=\half:[\oP^2+m^2\oQ^2]:$ is well defined and corresponds to the classical free Hamiltonian $H(\op,\oq)=\half[\op^2+m^2\oq^2]$. This
               assignment is valid for $N=\infty$ as well. The fiducial vector $|0\>$ is also the ground state for $\H_0$,
               namely, $\H_0\s|0\>=0$; in a Schr\"odinger representation, $\<\ox|0\>=N'\s\exp\{-(m/2\hbar)\s\ox^2\}$. In addition, the operator
                 $ \H_1\equiv \H_0+:\H_0^2: $
                 is also well defined for $N=\infty$, and with $|\op,\oq\>=U[\op,\oq]\s|0\>$, it follows that
                    \b H_1(\op,\oq)\equiv \<\op,\oq|\H_1|\op,\oq\>=\half[\op^2+m^2\oq^2]+\{\half[\op^2+m^2\oq^2]\}^2\;. \e
                    This expression contains $(\oq^2)^2$, but it also contains unwanted terms.

                    So far we have assumed the representation
                    for the operators $\oP$ and $\oQ$ is {\it irreducible}, but the classical/quantum relation
                    $H(\op,\oq)\equiv \<\op,\oq|\H|\op,\oq\>$ does {\it not} require that assumption.
                    Let us relax our assumptions and let the basic operator representation be {\it reducible}
                    but require that the set of coherent states $\{|\op,\oq\>\}$ span the abstract
                    Hilbert space.
                    Besides the set of operators $\oP$ and $\oQ$, we introduce a second, independent set of operators, namely, $\oR$ and $\oS$ with the commutation relation $[R_j,S_k]=i\hbar\delta_{j,k}\one$. The new Hamiltonian is taken to be
                    \b &&\H_2\equiv\half:[\oP^2+m^2(\oQ+\zeta\s\oS)^2]:+\half:[\oR^2+m^2(\oS+\zeta\s\oQ)^2]:\\
                    &&\hskip14.4em+\beta:[\oR^2+m^2(\oS+\zeta\s\oQ)^2]^2:\;, \e
                    where $\zeta$ is a new parameter with $0\le\zeta<1$. Each of the components, i.e,
                    $:[\oP^2+m^2(\oQ+\zeta\s\oS)^2]:$ and $:[\oR^2+m^2(\oS+\zeta\s\oQ)^2]:$, is like a
                    free field much as was $:[\oP^2+m^2\oQ^2]:$. Just as the vector $|0\>$ is the ground state of the latter operator, the vector $|0,0\>$ is the ground state for both of the former operators; in a Schr\"odinger representation, $\<\ox,\oy|0,0\>=N\s\exp\{-(m/2\hbar)[\ox^2+\oy^2+2\zeta \ox\cdot\oy]\}.$ The coherent states of interest are given by
                    $|\op,\oq\>\equiv U[\op,\oq]\s|0,0\>$; in a Schr\"odinger representation we have
                    \b&&\hskip-2em\<\ox,\oy|\op,\oq\>\\
                      &&\hskip-1.4em=N\s\exp\{i\op\cdot(\ox-\oq)/\hbar-(m/2\hbar)[(\ox-\oq)^2+\oy^2
                    +2\zeta(\ox-\oq)\cdot\oy]\}\,.\e
                     For $\zeta>0$, the functions $\<\ox,\oy|\op,\oq\>$ span
                    the space $L^2({\mathbb R}^2)$ of functions $\phi(\ox,\oy)$; however, if we set $\oq=0$, then the functions $\<\ox,\oy|\op,0\>$ do {\it not} span the same space. In this sense, the reducible representation of the operators spans a larger space not available to the irreducible representations. To complete the story we observe that
                     \b H_2(\op,\oq)\hskip-1.3em&&\equiv\<\op,\oq|\s\H_2\s|\op,\oq\>
                        =\half[\op^2+m^2(1+\zeta^2)\s\oq^2]+\beta\s m^4\zeta^4\s(\oq^2)^2\\
                        &&\equiv\half[\op^2+m^2_0\s\oq^2]+\l_0\s(\oq^2)^2\;. \e
                        Note that this desirable classical/quantum connection requires that $\zeta>0$ in order
                        to include the nonlinear interaction.

                        With this last expression, we have demonstrated a formulation of the quantum theory that
                        solves the Rotationally Symmetric models. When the number of degrees of freedom $N=\infty$, the free theory is the only case covered by irreducible representations
                        of the basic operators. To treat the interacting cases when $N=\infty$, it is essential to use reducible representations, and as a consequence the correct solution for interacting models
                        requires a violation of the usual dogma (of an irreducible representation).

                        This concludes our three stories
                        on how coherent states can help clarify the classical/quantum connection.
                        For additional stories see \cite{kla6,kla7}.

                        \end {document}